\documentclass[10pt]{article}
\usepackage[usenames,dvipsnames]{xcolor}
\usepackage[utf8]{inputenc}                   
\usepackage[pdftex]{graphicx} 
\usepackage{multicol} 
\usepackage[english]{babel}
\usepackage{float} 
\usepackage{amssymb} 
\usepackage{amsmath} 
\usepackage[pdftex]{hyperref}

\begin{document}
{\title{Study of cylindrically symmetric solutions in metric $f(R)$ gravity with constant $R$}
\author{M. T. Rincon-Ramirez \thanks{motrinconra@unal.edu.co - Physics Department, Universidad Nacional de Colombia}\\
L. Casta\~neda \thanks{lcastanedac@unal.edu.co - Observatorio Astronomico Nacional, Universidad Nacional de Colombia, Bogot\'a, Colombia}}
\date{}
\maketitle
\begin{abstract}
\noindent Solutions for cylindrically symmetric spacetimes in $f(R)$ gravity are studied. As a first approach, $R=$constant is assumed. A solution was found such that it is equivalent to a result given by Azadi et al. for $R=0$ and a metric was found for $R=constant\neq0$. Comparison with the case of general relativity with cosmological constant is made and the metric constants are given in terms of $\Lambda$.
\end{abstract}
\begin{multicols}{2}
\section{\label{sec:level1}Introduction}
\noindent Although general relativity is one of the most successful theories in physics, with countless verified predictions, it  took only four years from its release for people, like Weyl and Eddington, to question its uniqueness among gravitational theories and to propose adding higher order invariants in its action. In the 1960's, motivations from quantum field theory appeared. It was shown that general relativity is not renormalizable, so it can not be conventionally quantized, while its modifications with higher order terms in the Hilbert-Einstein action can  \cite{Sotiriou:2008rp}.
\\

\noindent The Nobel Prize in Physics 2011 was awarded to S. Perlmutter, B. P. Schmidt and A. G. Riess \textit{``for the discovery of the accelerating expansion of the Universe through observations of distant supernovae"} \cite{nobel}. If we take general relativity to be the ultimate gravitational theory, it turns out that our universe is made of approximately 4$\%$ of ordinary baryonic matter, 20$\%$ of dark matter and 76$\%$ of dark energy, that is characterized by a negative pressure and would be the cause of the accelerated expansion. This situation is disturbing for two main reasons: first, the two most abundant components of the universe have only been inferred by their gravitational interaction on cosmological scales (Dark matter on the scale of galaxies, clusters and the Hubble scale, and dark energy only on the Hubble scale), and second, considering particle physics, the cosmological constant is not distinguishable from the vacuum energy     \cite{Durrer:2011gq}, but the values from cosmology and quantum field theories differ in about 60 to 120 orders of magnitude.
\\

\noindent Nowadays, there is not a satisfactory explanation for the origin of dark energy. Among the multiple options considered, $f(R)$ gravity is a natural gravitational alternative; in $f(R)$ gravity the Einstein-Hilbert action is generalized, so that sub-dominant terms, relevant at small curvature, may explain today's accelerated expansion while avoiding dark components  \cite{Nojiri:2006ri}. There exist metric $f(R)$ gravity models, such as the Starobinsky model 
\begin{equation}
f(R)=R+\lambda R_0\left[ \dfrac{1}{\left(1+\frac{R^2}{R_0^2}\right)^n} -1 \right], 
\end{equation}
which have passed all the observational and theoretical constraints \cite{Capozziello:2009nq}. Additionally, dark-matter content of clusters have been proved to be satisfactory addressed by modified gravitational potential terms on distance scales from 150 kpc up to 1000 kpc by a direct fitting with data \cite{Capozziello:2008ny}.\\
 
\noindent In this sense, the finding of solutions in $f(R)$ gravity for highly symmetric spacetimes already studied in general relativity is a first step of crucial importance for the construction of qualified models with competent experimental predictions. Given the complexity of the field equations in $f(R)$, some assumption has to be made in order to simplify the equations and obtain analytical solutions, for example, imposing $R=$constant or specifying the functional form of $f(R)$. \\

\noindent Several breakthroughs have been made in this direction; in \cite{Hendi:2012zg}, S.H. Hendi et al. found analytic solutions with a nontrivial Eguchi-Hanson metric; S. H. Hendi et al. \cite{Hendi:2011eg} obtained an uncharged solution which corresponds with the topological (a)dS Schwarzschild solution and a solution with electric charge equivalent to the Einstein-$\Lambda$-conformally invariant Maxwell solution; for $f(R)=R^N$ S. H. Hendi \cite{Hendi:2009sw} found a charged black hole solution different from the standard higher-dimensional Reissner-Nordstrom solutions; in \cite{Hendi:2012nj} S. H. Hendi et al. obtained uncharged and charged solutions for some classes of $f(R)$ models with conformal anomaly; a class of higher dimensional rotating solutions in R+f(R) theory coupled to a conformally invariant Maxwell field is presented by A. Sheikhi et al. in \cite{sheykhi}, these solutions have asymptotically anti–de Sitter behavior and represent black branes with cylindrical or toroidal horizons; among others.  \\

\noindent Specifically, cylindrically symmetric and stationary spacetimes can be obtained as special cases of axisymmetric fields \cite{exactsol} and are of special interest in gravitation because of several reasons. For example, its relation with the topic of cosmic strings, which are a prediction of some unified theories of particle interactions and may help to explain some of the largest-scale structures seen in the Universe \cite{Hindmarsh:1994re}. The main interest in stationary axisymmetric vacuum solutions lies on the fact that they may describe the exterior gravitational field of rotating bodies, and of course on the abundance of such objects in our universe in the form of stars and dust. Azadi et al. have obtained cylindrically symmetric solutions for constant R and arbitrary f(R) in \cite{Azadi:2008qu}, that work is extended in \cite{Momeni:2009tk} by D. Momeni et al., where the metric is not limited to any specific gauge and the constant curvature solution is shown to be applicable to the exterior of a string. Under the same assumptions of \cite{Azadi:2008qu}, here we obtain a different solution with cilindrical symmetry for constant $R\neq$ 0.\\

\subsection{\label{1.2} Field equations and $R=0$}
\noindent Starting from the general cylindrically symmetric and stationary metric in simplified Weyl coordinates\cite{exactsol}$^{,}$ \cite{Azadi:2008qu}
\begin{multline}\label{metric}
ds^2=-e^{2k(r)-2u(r)}dt^2 + e^{2k(r)-2u(r)}dr^2 \\
+ w(r)^2e^{-2u(r)}d\phi^2 + e^{2u(r)}dz^2,
\end{multline}
\noindent the Ricci tensor's nonzero components for this metric are 
\begin{subequations}\label{RicciTensor}
\begin{align}
R_{tt}&= -k'' +u'' -\frac{w'k'}{w} + \frac{w'u'}{w}&\\
R_{rr}&=\frac{w''}{w} + k'' - u'' -\frac{w'k'}{w} + 2u'^2-\frac{u'w'}{w}&\\
R_{\phi\phi}&=-\frac{w}{e^{2k}}(u'w'-w''+wu'')&\\
R_{zz}&= e^{4u-2k}(u''+\frac{w'u'}{w}) &
\end{align}
\end{subequations}
\noindent where $ '=d/dr$. Therefore, the Ricci scalar is
\begin{eqnarray}\label{RicciScalar}
R=\frac{2}{we^{2k(r) -2u(r) }}&(-w(r)u''(r)+w(r)k''(r)\\
&-w'(r)u'(r)+w''(r)+w(r)u'(r)^2) \nonumber
\end{eqnarray}
\noindent which differs from the one obtained by Azadi et al.\cite{Azadi:2008qu} for the same metric.
\\

\noindent We focus on the study of cylindrically symmetric solutions in modified $f(R)$ gravity. In this approach the Einstein-Hilbert action, with Lagrangian $\mathcal{L}=R-2\Lambda$, is generalized to an action with $\mathcal{L}=f(R)$, where $f(R)$ is an arbitrary function of the Ricci scalar. The vacuum ($T_{\mu\nu}=0$) field equations in $f(R)$ gravity in the metric formalism are \cite{Guarnizo:2010xr}
\begin{equation}\label{FEfRvacuum}
F(R)R_{\mu\nu}-\frac{1}{2}f(R)g_{\mu\nu}-\nabla_\mu \nabla_\nu F(R) + g_{\mu\nu}\square F(R)=0
\end{equation}
with $F(R) \equiv df(R) /dR$ and $\square=\nabla_\mu\nabla^\mu$. \\
The contraction of the last equation,
\begin{equation}
F(R)R- 2f(R) + 3\square F(R)=0\label{FEtrace},
\end{equation}
is wisely used in \textit{Cylindrical solutions in metric $f(R)$ gravity\cite{Azadi:2008qu}} to leave the field equations only in terms of $F(R)$, instead of $f(R)$ and $F(R)$ together. Therefore, from (\ref{FEtrace}) and (\ref{FEfRvacuum}) it results
\begin{equation}
FR_{\mu\nu}-\nabla_\mu\nabla_\nu F\;= \;\frac{1}{4}g_{\mu\nu}(FR-\square F)\label{8}
\end{equation}
\noindent For the chosen diagonal metric, which depends only on $r$, equation (\ref{8}) constitutes a set of four differential equations for the functions $F(r)$, $u(r)$, $k(r)$ and $w(r)$. From the same Eq. (\ref{8}), one can define for \textbf{fixed} index $\mu$ 
\begin{equation}
A_\mu \equiv \frac{(FR_{\mu\mu}- \nabla_\mu \nabla_\mu F)}{g_{\mu\mu}}=\frac{1}{4}(FR-\square F(R))
\end{equation}
\noindent Since the right-hand side of this equation is independent of the index, $A_\mu =A_\nu$ $\forall$ $\mu,\nu$, we can form a set of three independent equations $A_t =A_r$, $A_r =A_\phi$ and $A_r =A_z$, which are shown below
\begin{subequations}
\begin{align}
F''-2F'(k'-u')-F(\frac{-2w'k'}{w}+\frac{w''}{w}+2u'^2)=0 \label{FE1}\\
Fw^2(-k''-\frac{w'k'}{w}+\frac{w''}{w})
+F'(k'w^2-w'w)=0&\label{FE2}\\
Fw(-k''+2u''-\frac{w'k'}{w}+ 2\frac{w'u'}{w})+F'(wk'-2wu')=0\label{FE3}
\end{align}
\end{subequations}
\noindent For the case $R=$constant, the above field equations are quite simplified to
\begin{subequations}
\begin{align}
\frac{-2w'k'}{w}+\frac{w''}{w}+2u'^2=0 \label{fe1}\\
-k''-\frac{w'k'}{w}+\frac{w''}{w}=0&\label{fe2}\\
-k''+2u''-\frac{w'k'}{w}+ 2\frac{w'u'}{w}=0\label{fe3}
\end{align}
\end{subequations}
or equivalently\cite{Azadi:2008qu}
\begin{eqnarray}
\frac{1}{2}\left(\frac{w'+c_2}{w}\right)^2 + \frac{w''}{w} =2\frac{w'}{w}\frac{c_1+w'}{w}\label{w'}\\
k'=\frac{c_1 + w'}{w}\label{k'}\\
u'=\frac{c_2+w'}{2w}\label{u'}
\end{eqnarray}
where $c_1$ and $c_2$ are integration constants.

\section{\label{2}Calculation with $u(r)=m k(r)$}
\noindent By imposing the condition $u(r)=m k(r)$, equations (\ref{fe1}),(\ref{fe2}),(\ref{fe3}) are
\begin{eqnarray}
w''-2w'k'+2m^2wk'^2=0 \label{m1}\\
w'k'-w''+wk'' =0  \label{m2}\\
(2m-1)(k''w+k'w')=0 \label{m3}
\end{eqnarray}
\noindent Equation (\ref{m3})shows that for $u(r)$ and $k(r)$ linearly dependent of each other, there are only two possibilities: $m=1/2$ or $k''w+k'w'=0$. The former leads to $R=constant\neq 0$ and the latter leads to $R=0\;\forall\; m$, as it will be illustrated bellow. \\

\noindent Considering only equations (\ref{m2}),(\ref{m3}) it is straightforward the solution
\begin{eqnarray}\label{generalsolution}
w(r)&=&d_1 r + d_2\\
u(r)/m &=& k(r)= d_4+d_3 \frac{\ln(r d_1 + d_2)}{d_1}
\end{eqnarray}
\noindent the last solution also satisfies the third field equation, Eq. (\ref{m1}), in two cases: a) $d_3$= 0, which is easily seen to correspond to Minkowski's metric, and b) $d_3 = d_1/m^2$ which brings
\noindent \begin{eqnarray}
w(r)&=&d_1 r + d_2\\
u(r)/m &=& k(r)= d_4+\ln(r d_1 + d_2)/m^2
\end{eqnarray}
\noindent we imposed this solution to satisfy $R=$constant, and the calculation of $R$ results in $R=0$ $\forall$ m.
\noindent The metric takes the form 
\begin{multline}\label{metricanueva}
ds^2= -D_4^{2(1-m)}\rho^{2\frac{1-m}{m^2}} dt^2 + \frac{D_4^{2(1-m)}}{d_1^2}\rho^{2\frac{1-m}{m^2}}d\rho^2 \\
 +  D_4^{-2m}\rho^{2\frac{m-1}{m}} d\phi^2  + D_4^{2m} \rho^{\frac{2}{m}} dz^2)
\end{multline}

\noindent with $D_4=e^{d_4}$ and $\rho=w(r)$. This solution is certainly non-minkowskian $\forall$ m$\neq 1$ and has definite values $\forall$ m$\neq 0$.
\\

\noindent More explicitly, the non-zero Riemann tensor's components for this metric are
\begin{subequations}
\begin{align}
&R_{t\rho t\rho}=D_4^{2(1-m)}\frac{(m-1)}{m^2}\rho^{-2\frac{m^2+m-1}{m^2}}\\
&R_{t\phi t\phi}=-d_1^2D_4^{-2m}\frac{(m-1)^2}{m^3}\rho^{-\frac{2}{m}}\\
&R_{tztz}=-d_1^2D_4^{2m}\frac{(m-1)}{m^3}\rho^{\frac{2(1-m)}{m}}\\
&R_{\rho\phi\rho\phi}= D_4^{-2m}\frac{(m-1)}{m^3}\rho^{\frac{-2}{m}}\\
&R_{\rho z\rho z}= D_4^{2m}\frac{(m-1)^2}{m^3}\rho^{\frac{2(1-m)}{m}}\\
&R_{\phi z\phi z}= -d_1^2 D_4^{2(m-1)}\frac{(m-1)}{m^2}\rho^{\frac{2(m-1)}{m^2}}
\end{align}
\end{subequations}
\noindent It should be noted that for $m=0$ the solution is not well-defined and for $m=1$ a change of coordinates leads to Minkowski's metric, otherwise, $m\neq1$ the Riemann tensor is not trivial and therefore it is a valid solution for cylindrically symmetric spacetimes. The constants $ d_1,\;D_4\;,m$ should be determined by boundary conditions, e.g. a mass distribution along the axis $r=0$.
\\

\noindent The above solution, obtained by setting $u(r)=mk(r)$, is equivalent to the metric obtained by Azadi et al. for $R=0$ by assuming $w(r)=c_7r$ from the beginning.\\

\noindent Even though substituting $m=1/2$ in the last result brings a solution with $R=0$, a different procedure can be carried on with $u(r)=\frac{1}{2}k(r)$, such that $w''(r)$ is not necessarily null, and leads to a different value of $R$ \cite{Azadi:2008qu}. Therefore, it can be concluded that both of them are particular independent solutions of the set of equations.

\section{ Solutions with $R=$constant$\neq0$}
\subsection{$k(r)=2u(r)$}
\noindent Assuming $k(r)=2u(r)$, equations (\ref{m1}), (\ref{m2}) and (\ref{m3}) are\cite{Azadi:2008qu}
\begin{eqnarray}
2u'^ 2+ \frac{w''}{w}-4\frac{u'w'}{w}=0 \label{n11} \\
2u''+\frac{2u'w'}{w}-\frac{w''}{w}=0  \label{n21} \\
0=0\label{n3}
\end{eqnarray}
using (\ref{n11}), (\ref{n21}) it is obtained
\begin{equation}\label{xx}
u''+u'^2-\frac{u'w'}{w}=0
\end{equation}
and the Ricci scalar (\ref{RicciScalar}) is
\begin{equation}\label{yy}
R\equiv R_0=2\frac{e^{-2u}w''}{w} 
\end{equation}
the solution for (\ref{xx}) and (\ref{yy}) is
\begin{eqnarray}
w(r)=-\frac{1}{r^2}\\
k(r)=\ln(\frac{12}{R_0 r^2})\\
u(r)=\frac{1}{2}\ln(\frac{12}{R_0 r^2})
\end{eqnarray}
by setting $R_0=4\Lambda$, the metric takes the form

\begin{equation}
\boxed{ds^2=\frac{3}{\Lambda r^2}\left(-dt^2+dr^2+\frac{\Lambda^2}{9}d\phi^2 +dz^2\right)}
\end{equation}

\noindent which clearly differs from that obtained in \textit{Cylindrically symmetric solutions in metric f(R) gravity}\cite{Azadi:2008qu} for $k(r)=2u(r)$. Such discrepancy is due to the difference in the Ricci scalar.\\

\subsection{Direct calculation}
\noindent Another approach consists on performing direct integration of the field equations (\ref{w'}), (\ref{u'}) and (\ref{k'}). It is defined $p\equiv p(w)=\frac{dw}{dr}$, in terms of which, equation (\ref{w'}) becomes
 
\begin{equation}\label{preintegral}
2wp'-3p^2+p(2c_2-4c_1) +c_2^2=0
\end{equation}

\noindent since $p'=dp/dr$ and $dr=dw(dr/dw)=dw/p$
\begin{subequations}\label{preintegral2}
\begin{align}
\frac{dp/dr}{-3p^2+p(2c_2-4c_1) +c_2^2}+\frac{1}{2w}=0\\
\int\frac{dp\;2p }{3p^2-p(2c_2-4c_1) -c_2^2}=\ln(w) + C \label{put}
\end{align}
\end{subequations}
The solution to this integral is 
\begin{eqnarray}
3I=&\frac{(2 c_1-c_2) \mathrm{arctanh}[\frac{2 c_1-c_2+3 p}{2 \sqrt{c_1^2-c_1 c_2+c_2^2}}]}{\sqrt{c_1^2-c_1 c_2+c_2^2}}\nonumber \\
&+\ln\left[4 c_1 p-(c_2-p) (c_2+3 p)\right]
\end{eqnarray}
\noindent which differs from that obtained in [\cite{Azadi:2008qu}] by a sign. By doing $c_2=2c_1$
\begin{equation}\label{paraw}
3p^2=(Aw)^3+4c_1^2 ,\qquad \mathrm{A\equiv e^C}
\end{equation}
\noindent from Eqs. (\ref{k'}), (\ref{u'}) and the definition of $p(w)$
\begin{subequations}
\begin{align}
r=\int\frac{dw}{p(w)}\label{intwr}&\\
u(w)=\frac{1}{2}\int\frac{p(w)+c_2}{wp(w)}dw \label{x} & \\
k(w)=\int\frac{p(w)+c_1}{wp(w)}dw \label{y}&
\end{align}
\end{subequations}
\noindent Computational methods lead to a complex equation for $w(r)$ and a solution for it can be given in terms of Weierstrass functions
\begin{equation}
w(r) = \mathtt{P}\left(\frac{2^{1/3}\sqrt{3}Ar}{6}+c_1; 0, -c^2\right)\frac{2^{2/3}}{A}
\end{equation}
Where $A$, $c$ and $c_1$ are constants.\\
Finally, concerning the comparison with the cosmological constant case, in spite of the differences in the Ricci scalar, the solution obtained agrees with that of [\cite{Azadi:2008qu}], being the Ricci scalar
\begin{equation}
R_0=A^3
\end{equation}

\noindent The obtainment of analytical solutions with $R(r)\neq constant$ will be postponed for future work.

\section{Conclusions}
\noindent From a general cylindrically symmetric metric in Weyl coordinates, a solution for the field equations in metric $f(R)$ was obtained by relating the metric functions in the form $u(r)=mk(r)$, such that it is equivalent to the metric obtained by Azadi et al. for $R=0$ by assuming $w(r)=c_7r$.\\

\noindent A solution for $R=constant\neq 0$ was obtained and it was compared to the cosmological constant case in general relativity. This solution differs from the calculation with $k(r)=2u(r)$ of Azadi et al. due to discrepancies in the Ricci scalar.

\section{Perspectives}
\noindent Apart from the study of more complex cases with $R\neq$ constant, gravitational lensing effects in axisymmetric spacetimes is another path to follow within the interest of the authors.

\end{multicols}
\end{document}